\documentclass[epsfig,12pt]{article}

\newcommand{\bce}{\begin{center}} 
\newcommand{\ece}{\end{center}}
\newcommand{\beq}{\begin{equation}}
\newcommand{\eeq}{\end{equation}}
\newcommand{\bea}{\vspace{0.25cm}\begin{eqnarray}}
\newcommand{\eea}{\end{eqnarray}}

\newcommand{\brho}{\mbox{\boldmath $\rho$}}

\newcommand{\bk}{{\bf k}}

\newcommand{\bp}{{\bf p}}
\newcommand{\bq}{{\bf q}}
\newcommand{\br}{{\bf r}}

\newcommand{\ba}{\begin{array}}
\newcommand{\ea}{\end{array}}

\newcommand{\ket}[1]{| {#1} \rangle}
\newcommand{\bra}[1]{\langle {#1} |}

\newcommand{\bfpsi}{\mbox{\boldmath $\psi$}}

\newcommand{\bkappa}{\mbox{\boldmath ${\kappa}$}}

\newcommand{\bDelta}{\mbox{\boldmath ${\Delta}$}}
\newcommand{\bb}{{\bf b}}
\def\lsim{\mathrel{\rlap{\lower4pt\hbox{\hskip1pt$\sim$}}
    \raise1pt\hbox{$<$}}}         %less than or approx. symbol
\def\gsim{\mathrel{\rlap{\lower4pt\hbox{\hskip1pt$\sim$}}
    \raise1pt\hbox{$>$}}}         %greater than or approx. symbol

\def\Pom{{\bf I\!P}}

\def\beq{\begin{equation}}
\def\endeq{\end{equation}}
\def\arr{\begin{eqnarray}}
\def\endarr{\end{eqnarray}}
\makeindex

  \advance\oddsidemargin  by -1in
  \advance\evensidemargin by -1in
  \advance\topmargin      by -0.5in
\begin{document}

\vspace{2.0cm}

\begin{flushright}
        FZJ-IKP-TH-2006-4       
\end{flushright}

\vspace{1.0cm}

\begin{center}
{\Large \bf 
Glue in the pomeron from nonlinear $k_\perp$-factorization}

\vspace{1.0cm}

{\large\bf N.N.~Nikolaev$^{a,b}$, W. Sch\"afer$^a$,
B.G.~Zakharov$^{b}$ and V.R.~Zoller$^{c}$}

\vspace{1.0cm}
{\sl
$^{a}$IKP(Theorie), FZ J{\"u}lich, D-52425 J{\"u}lich, Germany
\medskip\\
$^{b}$L.D. Landau Institute for Theoretical Physics, Moscow 117940, Russia
\medskip\\
$^{c}$ ITEP, Moscow 117218, Russia.
}\vspace{1.0cm}\\
{ \bf Abstract }\\
\end{center}

We derive the nonlinear $k_\perp$-factorization for the spectrum of jets
in high-mass diffractive deep inelastic scattering as a function of 
three hard scales - the 
virtuality of the photon $Q^2$, the transverse momentum of the jet and 
the saturation scale $Q_A$. In contrast to
all other hard reactions studied so far, we encounter a clash between 
the two definitions of the glue in the pomeron -- from the inclusive 
spectrum of leading quarks and the small-$\beta$ evolution of the diffractive 
cross section. This clash casts a further shadow on customary applications of 
the familiar collinear factorization to a pQCD analysis of diffractive 
deep inelastic scattering.

%\doublespace
\pagebreak

The high-mass, or leading log${1\over \beta}$ (LL${1\over \beta}$) 
evolution of Diffractive Deep Inelastic Scattering 
(DDIS) is subtle and remains under scrutiny 
(\cite{NZ92,NZ94,Bj,NZZdiffr,GNZ95,KolyaMC,NSSdijet,NSZZval,KovLev,Bartels,Golec,Goncalves,Marquet},
for the discussion in connection with the planned electron-ion
collider eRHIC see \cite{eRHIC}). 
When cast in the Good-Walker-Miettinen-Pumplin form
\cite{GoodWalker}, and upon the proper renormalization of wave functions,
the inclusive coherent forward DDIS
%looks as 
resembles a sum of differential cross sections
of elastic scattering of multiparton 
Fock states of the photon \cite{NZ92,NZ94}. As such, DDIS is an inherently 
nonlinear functional of the color dipole scattering 
amplitude to be contrasted to the linear one in inclusive DIS
\cite{NZ94,NZ91}. 
A customary treatment of DDIS as DIS off an isolated 
pomeron, viewed as a hadronic
state with the well defined flux in the target, an 
indiscriminate collinear factorization analysis of DDIS 
final states and conclusions on glue in the pomeron
from such an analysis (\cite{HERADDIS} and references therein) -- all 
must be taken with a grain of salt, especially in a regime  
dominated by multipomeron exchanges, as e.g. nuclear targets with
a large saturation scale. Here we present an analysis of 
glue in the pomeron as
inferred from the LL${1\over \beta}$ evolution of, and jets in, high-mass DDIS which 
strengthens these reservations. 

In the pQCD approach to DDIS, $\gamma^*p \to X+p'$ and $\gamma^*A \to X+A'$,  
high-mass DDIS starts with $X=q\bar{q}g$. Our  basic points derive 
from the comparison of 
nonlinear $k_\perp$-factorization for jets and LL${1\over \beta}$ evolution properties 
of
the $q\bar{q}g$ final states at very small 
DDIS Bjorken variable
$\beta = Q^2/(Q^2+M^2)\ll 1$, where $M$ is the invariant mass of the diffractive
state. As far as the LL${1\over \beta}$ evolution properties of DDIS are concerned,
one may insist on the linear color dipole  representation
(\cite{NZ94}, see also
the subsequent \cite{NSZZval,KovLev,Marquet}) similar to that
for inclusive DIS \cite{NZ91}. In 1993, such a color dipole form 
was derived within the double leading logarithm 
approximation (DLLA) and reinterpreted as a proof of the familiar DLLA 
evolution of DDIS \cite{NZ94}. Evidently, DLLA is inadequate for DDIS off nuclei 
(and nucleons) in the most interesting
regime of saturation. Furthermore, a utility of the emerging 
operational definition of the glue in
the pomeron, especially its  $k_\perp$-factorization and unitarity properties, 
remain unspecified. Indeed, the
fundamental point about inclusive DIS is that one and the same unintegrated glue
provides the linear $k_\perp$-factorization description of both the 
LL${1\over x}$ evolution and the leading quark spectra 
(\cite{Nonlinear} and references therein). 
Recently, this property has been shown to extend to hadron-nucleus collisions, too
\cite{ForwardLPM}. 
In this communication we show that for
high-mass DDIS the same property is not borne out by a rigorous nonlinear 
$k_\perp$-factorization for the final states and the evolution of the DDIS
cross section. Besides the identification of this clash, we report
a full analytic description of the spectra of leading jets in both
the photon and pomeron fragmentation regions of  DDIS as 
a function of three hard scales -- the virtuality of the photon $Q^2$,
the transverse momentum of the jet and the saturation scale $Q_A$ -- 
which extends earlier limited considerations 
\cite{KolyaMC,NSZZval,Golec}.

The first 
discussion of the $q\bar{q}g$ excitation goes back to the derivation of 
the color dipole BFKL equation \cite{NZ94,NZZBFKL}. In the  impact 
parameter space the excitation amplitude equals
$
{\cal A}(\bb,q\bar{q}g)=[\textsf{S}_{q\bar{q}g}(\bb;\brho,\br)-
\textsf{S}_{q\bar{q}}(\bb;\br)]\Psi_{q\bar{q}g}(\brho,\br)$, 
where $\Psi_{q\bar{q}g}(\brho,\br)$ is the lightcone wave function of the 
$q\bar{q}g$ state, $\br$ and
$\brho$ are the $q\bar{q}$ and $qg$ dipoles, $\textsf{S}_{q\bar{q}g}(\bb;\brho,\br)$ 
and $\textsf{S}_{q\bar{q}}(\bb;\br) $ are
the $\textsf{S}$-matrices of the $q\bar{q}g$ and $q\bar{q}$ Fock states at
the impact parameter $\bb$ (we follow the 
notations of \cite{NZ94,Nonlinear,SingleJet,GluonGluonDijet}). Inclusive 
forward DDIS off nucleons (N) evolves as
%, %which is free of uncertainties
%with the diffraction slope, 
\bea
16\pi\left.{\frac{d\sigma_N^D}{ dt}}\right|_{t=0}=\bra{ q\bar{q}}\sigma_{q\bar{q}}^2\ket{q\bar{q}}
+\bra{ q\bar{q}g}\sigma_{q\bar{q}g}^2-
\sigma_{q\bar{q}}^2\ket{q\bar{q}g}+... 
\label{eq:1}
\eea
For heavy nuclei (A) the integration over the momentum transfer
can be carried out explicitly and we consider the coherent 
DDIS cross section per unit area in the impact parameter space:
\bea
\frac{d\sigma_A^D}{d^2\bb}=
\bra{ q\bar{q}}|1-\textsf{S}_{q\bar{q}}|^2\ket{q\bar{q}}
+\bra{q\bar{q}g}|1-\textsf{S}_{q\bar{q}g}|^2-|1-\textsf{S}_{q\bar{q}}|^2\ket{q\bar{q}g}+....
\label{eq:2}
\eea
Apart from the excitation of the physical  $q\bar{q}g$ state, 
the $q\bar{q}g$ contribution to (\ref{eq:1}), (\ref{eq:2}) 
describes also the LL${1\over \beta}$ evolution of the $q\bar{q}$ excitation
\cite{NSZZval}.  For both targets, the
transverse momenta of jets in the $q\bar{q}g$ final state add
to zero: $\bp_q+\bp_{\bar{q}}+\bp_g=0$. Hereafter $\bp \equiv \bp_q$,
$\bq\equiv \bp_g$, the rapidity gap variable $x_\Pom=x/\beta$ is 
a fraction of nucleon's momentum carried by the pomeron,
$z$ is the Feynman variable -- a fraction of the photon's lightcone momentum carried
by the quark.

We derive  nonlinear $k_\perp$-factorization directly for the 
observed DDIS cross section
without the separation of the ill-defined flux of pomerons which is not
borne out by the pQCD treatment of DDIS \cite{GNZ95}. One of
the building blocks is the lightcone 
wave function of the $q\bar{q}$ state of the photon and $qg$ Fock 
state of the quark. For transverse photons, in terms of the 
QED splitting function
$P_{q\gamma}(z)=2N_c e_f^2\alpha_{em} [z^{2}+(1-z)^{2}]$ one has
$
|\Psi_{q\bar{q}}(z,\bp)-\Psi_{q\bar{q}}(z,\bp-\bkappa)|^2= P_{q\gamma}(z)
\left|\bfpsi(\varepsilon^2,\bp)-
\bfpsi(\varepsilon^2,\bp-\bkappa)\right|^2$,
where $\bfpsi(\varepsilon^2,\bp)=\bp/(\varepsilon^2+\bp^2)$  
and $\varepsilon^2=z(1-z)Q^2+m_f^2$ (we focus on
light flavors, a simple extension to longitudinal photons and heavy
flavors will be reported elsewhere). In high-mass DDIS gluons are soft, 
$z_g\ll 1$, and $\Psi_{qg}(z_g,\bp)$ does not
depend on the virtuality of the parent quark,
$
|\Psi_{qg}(z_g,\bp)-\Psi_{qg}(z_g,\bp-\bkappa)|^2=
4\alpha_S C_F \left|\bfpsi(\mu^2,\bp)-
\bfpsi(\mu^2,\bp-\bkappa)\right|^2/z_g$ ,
where the (optional) infrared parameter
$\mu$ models the finite propagation radius of perturbative gluons
\cite{NZ94,NZZBFKL}. The technique of 
Refs. \cite{Nonlinear,SingleJet,GluonGluonDijet} gives the 
nonlinear $k_\perp$-factorization
master formulas (for nuclei we cite the leading term of
the large-$N_c$ perturbation theory, where $N_c$ is the number of colors)
\bea
\left.{\partial \over \partial \log{1\over \beta}}{ d\sigma_N^D \over dz d^2\bp d^2\bq  dt}\right|_{t=0}
={\frac{1}{ 16\pi (2\pi)^4}} \cdot\left({C_A\over 2C_F}\right)^2\cdot
4\alpha_S C_F P_{q\gamma}(z) \left|\int d^2\bkappa f(x_\Pom,\bkappa) H_{ij}^N(\bkappa,\bq,\bp)\right|^2, 
\label{eq:3}
\eea
\bea
{\partial \over \partial \log{1\over \beta}}{d\sigma_A^D \over dz  d^2\bb d^2\bp d^2\bq }
&=&{\frac{1}{ (2\pi)^4}} 4\alpha_S C_FP_{q\gamma}(z) \nonumber\\
&\times& 
\left|\int d^2\bkappa_1 d^2\bkappa_2\Phi(\bb,x_\Pom,\bkappa_1) 
\Phi(\bb,x_\Pom,\bkappa_2) H_{ij}^A(\bkappa_1,\bkappa_2,\bq,\bp)\right|^2,
\label{eq:4}
\eea
where 
\bea
 H_{ij}^N(\bkappa,\bq,\bp)&=& 
\bfpsi_i(\mu^2,\bq)
\Big\{[\bfpsi_j(\varepsilon^2,\bp+\bq)-\bfpsi_j(\varepsilon^2,\bp)]%\nonumber\\
%&+&
+[\bfpsi_j(\varepsilon^2,\bp-\bkappa+\bq)-
\bfpsi_j(\varepsilon^2,\bp-\bkappa)]\Big\}
\nonumber\\
&-&\{\bq \to \bq+\bkappa\},
%\bfpsi_i(\mu^2,\bq+\bkappa)
%\Big\{[\bfpsi_j(\varepsilon^2,\bp+\bq+\bkappa)-
%\bfpsi_j(\varepsilon^2,\bp)]\nonumber\\
%&+&
%[\bfpsi_j(\varepsilon^2,\bp-\bkappa+(\bq+\bkappa))-
%\bfpsi_j(\varepsilon^2,\bp-\bkappa)]
%\Big\},
\label{eq:5}
\eea
\bea
&& H_{ij}^A(\bkappa_1,\bkappa_2,\bq,\bp) = 
\bfpsi_i(\mu^2,\bq)
[\bfpsi_j(\varepsilon^2,\bp-\bkappa_1+\bq)-\bfpsi_j(\varepsilon^2,\bp-\bkappa_1)]
-\{\bq \to \bq+\bkappa_1+\bkappa_2\},\nonumber
\\
%&-&\bfpsi_i(\mu^2,\bq+\bkappa_1+\bkappa_2)
%[\bfpsi_j(\varepsilon^2,\bp-\bkappa_1+(\bq+\bkappa_1+\bkappa_2))-
%\bfpsi_j(\varepsilon^2,\bp-\bkappa_1)]\Big\},%\nonumber
\label{eq:6}
\eea
the free-nucleon glue $f(x_\Pom,\bkappa)$ is related to the $q\bar{q}$
color dipole cross section by $\sigma(x,\br)=\int d^2\bkappa f(x,\bkappa)[1-\exp(i\bkappa\cdot \br)]$,
$\Phi(\bb,x_\Pom,\bkappa)= {\textsf S}[\bb;\sigma_0(x_\Pom)]
\delta^{(2)}(\bkappa)+\phi(\bb,x_\Pom,\bkappa)$ is defined in terms of the
nuclear ${\textsf S}$-matrix for the $q\bar{q}$ dipole \cite{Nonlinear,SingleJet,GluonGluonDijet},
${\textsf S}[\bb;\sigma_0]=
\exp[-{1\over 2}\sigma_0 T(\bb)]$, $T(\bb)$ is the optical thickness
of the nucleus and $ \sigma_0(x_\Pom)=\int d^2\bkappa f(x_\Pom,\bkappa)$. For heavy nuclei 
a useful analytic approximation is \cite{Nonlinear}
$
\Phi(\bb,x_\Pom,\bkappa) \approx 
Q_A^2(\bb,x_\Pom)/\pi (Q_A^2(\bb,x_\Pom) + \bq^2)^2,$
where $
Q_A^2(\bb,x)\approx {4\pi^2 \over N_c}\alpha_S(Q_A^2)G_N(x,Q_A^2)T(\bb) 
$
is the saturation scale \cite{Nonlinear} and $G_N(x,Q^2)$ is the integrated glue
of the free nucleon. For $\bq^2 \gg Q_A^2$ we shall often
encounter 
\beq
\int^{\bq^2} d^2\bkappa ~\bkappa^2\Phi(\bb,x_\Pom,\bkappa) 
\approx {1\over 2}Q_A^2(\bb,x_\Pom) {\alpha_S(\bq^2)G_N(x,\bq^2)
\over \alpha_S(Q_A^2)G_N(x,Q_A^2)}.
\label{eq:7}
\eeq

The jet-integrated DDIS cross section can be cast in the color dipole form,
\bea
{\partial \over \partial \log{1\over \beta}}\left.{\frac{d\sigma_{N}^{D}} { dt}}\right|_{t=0} = 
\langle q\bar{q}|\sigma_{N}^{\Pom}(x_{\Pom},\beta,\br)|q\bar{q}\rangle ,
\label{eq:8}
\eea
which defines the LL${1\over \beta}$ unintegrated glue in the pomeron via  $
\sigma_N^{\Pom}(x_{\Pom},\beta,\br)=\int d^2\bkappa
f^{\Pom}(x_{\Pom},\beta,\bkappa)\cdot [1-\exp(i\bkappa\cdot\br)]$ 
(a somewhat odd dimension, $[\sigma_{N}^{\Pom}]= [{\rm mb \cdot GeV^{-2}}]$ is
unimportant): 
\bea
&&f_N^{\Pom}(x_{\Pom},\beta,\bp)=
{1\over 16\pi} \cdot {C_A\over C_F}\cdot {C_A\alpha_S \over 2\pi^2}
\int d^2\bkappa_1 d^2\bkappa_2 \Biggl\{ f(x_{\Pom},\bkappa_1) f(x_{\Pom},\bkappa_2)\nonumber\\
&&\times 
\Big[4K(\bp,\bp+\bkappa_1)-2K(\bp,\bp+\bkappa_1+\bkappa_2) \nonumber \\
&&+4K(\bp+\bkappa_1,\bp+\bkappa_1+\bkappa_2)- 2K(\bp+\bkappa_1,\bp+\bkappa_2)\Big]
\nonumber\\
&&-  f(x_{\Pom},\bkappa_1) f(x_{\Pom},\bp- \bkappa_1)\Big[
2K(\bkappa_2,\bkappa_2+\bkappa_1)
%\nonumber\\
%&+&
%+K(\bkappa_2,\bp+\bkappa_2-\bkappa_1)
-K(\bkappa_2,\bp+\bkappa_2)\Big]
\nonumber\\
&&- 2f(x_{\Pom},\bp) f(x_{\Pom},\bkappa_1)\Big[
K(\bkappa_2,\bkappa_2+\bp)
%\nonumber\\
%&+&
+K(\bkappa_2,\bkappa_2-\bkappa_1)-
K(\bkappa_2,\bp+\bkappa_1+\bkappa_2)\Big]\Biggr\}\nonumber\\
%\label{eq:11}
&&=
{1\over 16\pi}{C_A\over 2C_F} {\cal K}_{BFKL} \otimes f_D(x_\Pom,\bp)% \nonumber\\
%&+& 
+{1\over 16\pi} \cdot {C_A\over C_F}{\cal K}_0
\int d^2\bkappa_1 d^2\bkappa_2\Biggl\{ f(x_{\Pom},\bkappa_1) f(x_{\Pom},\bkappa_2)
\nonumber\\
&&\times \Big[2K(\bp+\bkappa_1,\bp+\bkappa_1+\bkappa_2)-K(\bp+\bkappa_1,\bp+\bkappa_2)]
\nonumber\\
%&-&
&&- f(x_{\Pom},\bkappa_1) f(x_{\Pom},\bp- \bkappa_1)K(\bkappa_2,\bkappa_1+\bkappa_2)\nonumber\\
&&-f(x_{\Pom},\bp) f(x_{\Pom},\bkappa_1)\Big[
K(\bkappa_2,\bkappa_2+\bkappa_1)-
K(\bkappa_2,\bp+\bkappa_1+\bkappa_2)\Big]\Biggr\},
\label{eq:9}
\eea
where $K(\bp,\bq)=|\bfpsi(\mu^2,\bp)-\bfpsi(\mu^2,\bq)|^2$.
The first form of this nonlinear $k_\perp$-factorization
result looks as a fusion of two pomerons,
described by $f(x_{\Pom},\bkappa_{1,2})$, into the third one described
by $f_N^{\Pom}(x_{\Pom},\beta,\bp)$.
In the second form  we singled out the linear BFKL evolving component with
$f_D(x_\Pom,\bp)=\sigma_0(x_\Pom) [ 2f(x_\Pom,\bp)- f^{(2)}(x_\Pom,\bp)]$ which
 describes 
$\sigma^2(x_\Pom,\br)= \int d^2\bkappa f_D(x_\Pom,\bp)[1-\exp(i\bkappa\cdot \br)]$
\cite{NSSdijet}, where $f^{(2)}(x_\Pom,\bp)= {1\over \sigma_0(x_\Pom)}\Big( f\otimes f\Big)(x_\Pom,\bp)$
and 
${\cal K}_0 = C_A\alpha_S/2\pi^2$. The interpretation of $f_D(x_\Pom,\bp)$ as a gluon density
must be taken with a big grain of salt as it is not positive valued, see the
discussion of the (anti)shadowing properties of $f^{(j)}(x,\bp)$
in Ref. \cite{NSSdijet}. 
The related LL${1\over \beta}$ result for high-mass DDIS off heavy nuclei reads (here 
$\Phi^{(2)}=\Phi\otimes \Phi$)
\bea
\Phi_A^{\Pom}(\bb,x_{\Pom},\beta,\bp)&=& 
{\cal K}_0 \Biggl\{ \int d^2\bkappa_1  d^2\bkappa_2 d^2\bkappa_3
\nonumber\\
&\times& 
\Big[2\Phi(\bb,x_\Pom,\bkappa_1)\Phi(\bb,x_\Pom,\bkappa_2)\Phi(\bb,x_\Pom,\bkappa_3)
K(\bp-\bkappa_1-\bkappa_2,\bp-\bkappa_1-\bkappa_3)\nonumber\\
&-&2\Phi(\bb,x_\Pom,\bkappa_1)\Phi(\bb,x_\Pom,\bkappa_2)\Phi(\bb,x_\Pom,\bp-\bkappa_1) 
K(\bkappa_3,\bkappa_1+\bkappa_2+\bkappa_3)\Big]\nonumber\\
&+& \Phi^{(2)}(\bb,x_\Pom,\bp)\int d^2\bkappa_1  d^2\bkappa_2 
\Phi^{(2)}(\bb,x_\Pom,\bkappa_2)K(\bkappa_1,\bp+\bkappa_1+\bkappa_2)\nonumber\\
&-&\int d^2\bkappa_1  d^2\bkappa_2  \Phi^{(2)}(\bb,x_\Pom,\bkappa_1)
\Phi^{(2)}(\bb,x_\Pom,\bkappa_2)K(\bp-\bkappa_1,\bp-\bkappa_2)\Biggr\}.
\label{eq:10}
\eea

Whether the LL${1\over \beta}$ inspired operational definitions
(\ref{eq:9}) and (\ref{eq:10}) as unintegrated glue in the
pomeron are viable or not, can
only be decided upon inspecting their utility for the description of 
diffractive final states. To this end we 
recall a remarkable 
linear $k_{\perp}$-factorization for the forward $q\bar{q}$ dijets
in DIS off free nucleons 
\bea
&&\frac{2(2\pi)^2d\sigma_N(\gamma^*\to q\bar{q})}
% \over 
{dz d^2\bp  d^2\bDelta} =  f(x, \bDelta ) P_{q\gamma}(z)
 \left|\bfpsi(\varepsilon^2,\bp) -
\bfpsi(\varepsilon^2,\bp-\bDelta )\right|^2 \,.
\label{eq:11}
\eea
Here the distribution of the experimentally measurable
dijet acoplanarity momentum, $\bDelta=\bp_q + \bp_{\bar{q}}$,
is described by 
precisely the same unintegrated glue $ f(x, \bDelta )$ which
describes the  LL${1\over x}$ evolution of DIS.
In diffractive $q\bar{q}g$ final states $\bDelta=-\bq$. 
Following Ref.~\cite{Nonlinear} we observe that one can cast
the DDIS cross section (\ref{eq:8}) in the Fourier form,
\beq
\langle q\bar{q}|\sigma_N^{\Pom}(x_{\Pom},\beta,\br)|q\bar{q}\rangle=
 \int_0^1 dz P_{q\gamma}(z)\int d^2\bp d^2\bq \, \,  f_N^\Pom (x_\Pom,\beta,\bq ) 
 \left|\bfpsi(\varepsilon^2,\bp) -
\bfpsi(\varepsilon^2,\bp-\bq )\right|^2.
\label{eq:12}
\eeq
Undoing the $z,\bp$ and $\bq$ integrations in (\ref{eq:12}), and 
treating $\bp$ as the quark jet momentum, i.e.,
enforcing a certain unitarity interpretation on the mathematical
Fourier representation (\ref{eq:12}), one  would obtain
precisely the form (\ref{eq:11}) for the diffractive dijet spectrum.
However, such a reinterpretation does not match the dijet cross section
given by Eqs. (\ref{eq:3}), (\ref{eq:5}). The source of this clash is that
(\ref{eq:12}) derives from
the integral form of (\ref{eq:3}) upon shifts of the integration
variables $\bp,\bq$, and after such shifts their meaning of the 
observed jet momentum
is lost. Such shifts signal the distortions of dipoles by multipomeron
exchanges which is familiar from our nonlinear
$k_\perp$-factorization results for the single-jet and dijet spectra
\cite{Nonlinear,SingleJet,GluonGluonDijet} -- in DDIS such a nonlinearity
and the found mismatch between the two definitions of the glue in the 
pomeron persist already for the free-nucleon target. 
We conclude that the DDIS dijet spectrum, given by the 
correct unitarity cut (\ref{eq:3}), is not 
linear $k_\perp$-factorizable in terms of the pomeron glue
(\ref{eq:9}) operationally defined by the LL${1\over \beta}$
color dipole representation for 
DDIS. 

In inclusive DIS, the leading quark spectrum is linear $k_\perp$-factorizable
even for nuclear targets:
\bea
&&\frac{(2\pi)^2d\sigma_A(\gamma^*\to q\bar{q})}
% \over 
{dz d^2\bp  d^2\bb} =  P_{q\gamma}(z)\int d^2\bq \, \,  \phi(\bb,x,\bq) 
 \left|\bfpsi(\varepsilon^2,\bp) -
\bfpsi(\varepsilon^2,\bp-\bq )\right|^2 \,.
\label{eq:13}
\eea
Repeating the considerations around (\ref{eq:12}), we would 
obtain for DDIS dijets the representation (\ref{eq:13}) in terms
of $\phi_A^\Pom (\bb,x_\Pom,\beta,\bq )$, 
which must be compared to (\ref{eq:4}) integrated over the gluon
momentum $\bq$. Upon such a comparison, we reiterate 
the already made point:  the spectrum of leading
quarks in DDIS off nuclei is not linear $k_\perp$-factorizable in terms 
of the (nonlinear) LL${1\over \beta}$ 
evolution-defined glue in the nuclear pomeron, the
failure of the latter
to describe the diffractive finial states shows it is not an observable
with the appropriate unitarity cut properties. 

Now we turn to the experimentally measurable single-jet spectra. We 
go directly to the most
interesting case of nuclear targets with large saturation scale,
$Q_A^2(\bb,x_\Pom) \gg \mu^2$, and illustrate our technique with hard 
leading quark jets, $\bp^2 +\varepsilon^2 \gg Q_A^2(\bb,x_\Pom)$,
a detailed treatment for the free-nucleon target will be reported elsewhere.  
This spectrum  is dominated by the contribution
from $\bkappa_i^2,\bq^2 \lsim Q_A^2(\bb,x_\Pom)$,
and one can expand
\bea
 H_{ij}^A(\bkappa_1,\bkappa_2,\bq,\bp) = {1\over \bp^2 +\varepsilon^2}
B_{jk}\Big[{\bq_i \bq_k \over \bq^2} - {(\bq+\bkappa_1+\bkappa_2)_i
(\bq+\bkappa_1+\bkappa_2)_k \over (\bq+\bkappa_1+\bkappa_2)^2}\Big],
\label{eq:14}
\eea
with  $B_{jk} = \delta_{jk} - 2\bp_j \bp_k (\bp^2+\varepsilon^2)^{-1}.$
Now notice that, upon the azimuthal integrations 
\bea
&& \langle H_{ij}^A(\bp,\bq) \rangle \equiv 
\int d^2\bkappa_1 d^2\bkappa_2\Phi(\bb,x_\Pom,\bkappa_1) 
\Phi(\bb,x_\Pom,\bkappa_2) H_{ij}^A(\bkappa_1,\bkappa_2,\bk,\bp)\nonumber\\
%&=& B_{jk}\int d^2 \bkappa\Phi^{(2)}(\bb,x_\Pom,\bkappa) 
%\Big[{\bq_i \bq_k \over \bq^2} - {(\bq+\bkappa)_i
%(\bq+\bkappa)_k \over (\bq+\bkappa_1)^2}\Big]\nonumber\\
&=& B_{jk}
\left({\bq_i \bq_k \over \bq^2}-{1\over 2}\delta_{ik}\right)
\int d^2 \bkappa\Phi^{(2)}(\bb,x_\Pom,\bkappa)\left[
\theta(\bkappa^2-\bq^2) +{\bkappa^2\over \bq^2}\theta(\bq^2 -\bkappa^2)\right]\nonumber\\
&=&  B_{jk}\left({\bq_i \bq_k \over \bq^2}-{1\over 2}\delta_{ik}\right) 
C(2Q_A^2(\bb,x_\Pom),\bq^2),
\label{eq:15}
\eea
where
\bea
C(Q_A^2,\bq^2)\approx 
\left[{Q_A^2 \over Q_A^2 + \bq^2} + 
{Q_A^2 \over \bq^2 } \left(
\log{Q_A^2 + \bq^2\over Q_A^2 } - {\bq^2 \over 
Q_A^2 +\bq^2} \right)\right]
\label{eq:16}
\eea
vanishes at  $\bq^2 \gsim Q_A^2$ and, as it was anticipated,  $\int d^2\bq C^2(Q_A^2,\bq^2)$ converges 
at $\bq^2 \sim Q_A^2$.

Now we notice that to DLLA the integrand of (\ref{eq:13}) for inclusive DIS equals  
\beq
\phi_A(\bb,x,\bq) (\bq_i B_{ik})^2{1\over (\bp^2 +\varepsilon^2)^2} \approx
\phi_A (\bb,x,\bq){\bq^2 (\bp^2+\varepsilon^2)^2 - 
4\varepsilon^2 (\bp\cdot\bq)^2 \over (\bp^2 +\varepsilon^2)^4}
\label{eq:17}
\eeq
and one can try to make a contact with DDIS in which 
the similar r\^ole is played by 
\bea
 \langle H_{ij}^A(\bp,\bq) \rangle^2 = {\bp^4+\varepsilon^4
\over 2(\bp^2+\varepsilon^2)^4}C^2(2Q_A^2,\bq^2).
\label{eq:18}
\eea
The two integrands, (\ref{eq:17}) and (\ref{eq:18}), have different dependence on
$(\bp\cdot \bq)$. With this reservation, a comparison of the two
suggests 
\beq
\phi_A^\Pom (\bb,x_\Pom,\beta,\bq) \sim {1\over \bq^2} C^2(2Q_A^2,\bq^2),
\label{eq:19}
\eeq
which has a pure higher twist behavior 
$\phi_A^\Pom (\bb,x_\Pom,\beta,\bq) \sim 1/ \bq^6$ for $\bq^2 \gsim Q_A^2$,
to be contrasted with $\phi (\bb,x,\bq) \sim 1/ \bq^4$ in
inclusive DIS.

Omitting the technicalities of the derivation for minijets, $\bp^2\lsim Q_A^2$, 
we cite our principal result
\beq 
{\partial \over \partial \log{1\over \beta}}{d\sigma_A^D \over dz d^2\bb d^2\bp } \sim
{Q_A^2(\bb,x_\Pom) \over [Q_A^2(\bb,x_\Pom) +\varepsilon^2 +\bp^2]^2},
\label{eq:20}
\eeq
which interpolates between all regimes and is
one of the novelties of our paper.  
For hard jets, $\bp^2 +\varepsilon^2 \gg Q_A^2(\bb,x_\Pom)$, the
expansion (\ref{eq:14}) amounts to vanishing intranuclear
distortions. Indeed, in this limit the
impact parameter integration gives
$d\sigma_A^D  \propto \int d^2\bb Q_A^2(\bb,x_\Pom) \propto A^1 d\sigma_N^D$,
while in the opposite limit of minijets, $\bp^2 +\varepsilon^2 \ll Q_A^2(\bb,x_\Pom)$,
there is an obvious strong nuclear suppression (cf. the discussion in \cite{SingleJet}), 
\beq
R_{A/N} = {d\sigma_A^D\over A d\sigma_N^D} \propto  
\left({\bp^2 +\varepsilon^2 \over  Q_A^2(\bb,x_\Pom)}\right)^2
\sim A^{-2/3}.
\label{eq:21}
\eeq
We emphasize the importance of measuring leading quark jets at fixed $Q^2$
as a function of their Feynman variable $z$, as that would give a
handle on $\varepsilon^2$ and the width of the plateau,
$Q_A^2(\bb,x_\Pom) +\varepsilon^2$, and 
allow an accurate determination of the
saturation scale $Q_A^2$.
 
Now we turn to leading jets in the pomeron fragmentation
region, i.e., gluons at the boundary of the rapidity gap. Here one
integrates over the quark momenta with the result
\bea
&&\int d^2\bp \left|\int d^2\bkappa_1 d^2\bkappa_2\Phi(\bb,x_\Pom,\bkappa_1) 
\Phi(\bb,x_\Pom,\bkappa_2) H_{ij}^A(\bkappa_1,\bkappa_2,\bq,\bp)\right|^2\nonumber\\
%&=& \int d^2\bkappa_1 d^2\bkappa_2 d^2\bkappa_3 d^2\bkappa_4 
%\Phi(\bb,x_\Pom,\bkappa_1) 
%\Phi(\bb,x_\Pom,\bkappa_2)\Phi(\bb,x_\Pom,\bkappa_3) 
%\Phi(\bb,x_\Pom,\bkappa_4)\nonumber\\
%&\times& 
%\Big[\bfpsi_i(\mu^2,\bq)\bfpsi_i(\mu^2,\bq)-
%2\bfpsi_i(\mu^2,\bq+\bkappa_1+\bkappa_2)\bfpsi_i(\mu^2,\bq)\nonumber\\
%&+&\bfpsi_i(\mu^2,\bq+\bkappa_1+\bkappa_2)\bfpsi_i(\mu^2,\bq+\bkappa_3+\bkappa_4)\Big]
%\nonumber\\
%&\times&
%\Big[\Omega(\bq+\bkappa_1+\bkappa_3) - \Omega(\bkappa_1-\bkappa_3)\Big],\nonumber\\
&=&\int d^2\bkappa_1 d^2\bkappa_2 \Phi(\bb,x_\Pom,\bkappa_1)%\nonumber\\
%&\times&
 \Phi(\bb,x_\Pom,\bkappa_2)\Big[\bfpsi_i(\mu^2,\bq)\bfpsi_i(\mu^2,\bq)-
2\bfpsi_i(Q_A^2,\bq+\bkappa_1)\bfpsi_i(\mu^2,\bq)\nonumber\\
&+&\bfpsi_i(Q_A^2,\bq+\bkappa_1)\bfpsi_i(Q_A^2,\bq+\bkappa_2)\Big]
%\nonumber\\
%&\times&
\cdot\Big[\Omega((\bq+\bkappa_1+\bkappa_2)^2) - \Omega((\bkappa_1-\bkappa_2)^2)\Big],
\label{eq:22}
\eea
where we used $
\int d^2\bkappa \Phi(\bb,x_\Pom,\bkappa) \bfpsi(\mu^2,\bp-\bkappa)\approx
\bfpsi(Q_A^2,\bp)$ valid for the interesting case of heavy nuclei with 
$Q_A^2 \gg\mu^2$. 
The dependence on the virtuality of the photon $Q^2$ is concentrated in 
($a^2=\bq^2/2\varepsilon^2$)
\bea
\Omega(\bq^2)=\int d^2\bp |\bfpsi(\varepsilon^2,\bp+\bq)-
\bfpsi(\varepsilon^2,\bp)|^2%\nonumber\\
%&=& 
=2\pi\Biggl({1+2a^2 \over 2a \sqrt{1+a^2}} 
\log { \sqrt{1+a^2}+a\over\sqrt{1+a^2}-a} -1 \Biggr), 
\label{eq:23}
\eea

For hard gluon jets,
$\bq^2\gg Q_A^2$, the dominant contribution would come
from $\bkappa_i^2 \ll \bq^2$, when one can expand
$\Omega((\bq+\bkappa_1+\bkappa_2)^2) \approx \Omega(\bq^2)
+ 2\Omega'(\bq^2)(\bq\cdot \bkappa_1 +\bq\cdot \bkappa_2)$. The first term of this
expansion gives a rise to a contribution to (\ref{eq:22}) of the form
\bea
&&\Omega(\bq^2)\left\{\int d^2\bkappa \Phi(\bb,x_\Pom,\bkappa)
\left[\bfpsi(\mu^2,\bq)- \bfpsi(Q_A^2,\bq+\bkappa)\right]\right\}^2 =
\nonumber\\
&\approx& { \Omega(\bq^2) \over \bq^2}\cdot\left({Q_A^2\over Q_A^2 + \bq^2}
\right)^2 \cdot \left[{\alpha_S(\bq^2+Q_A^2)G_N(x,\bq^2+Q_A^2)
\over \alpha_S(Q_A^2)G_N(x,Q_A^2)}\right]^2.
\label{eq:24}
\eea
Making use of Eq. (\ref{eq:23}), one can readily check that the contribution
from the second term, 
$\Omega'(\bq^2)(\bq\cdot \bkappa_{1,2})$, will have a similar large-$\bq^2$ asymptotics. 

For soft gluons, $\bq^2 \ll \bkappa_{1,2}^2 \sim Q_A^2$, there is
an apparent small-$\bq$ singularity of $\bfpsi(\mu^2,\bq)$.
Upon the expansion, $\Omega((\bq+\bkappa_1+\bkappa_2)^2) - \Omega((\bkappa_1-\bkappa_2)^2)
\approx \Omega'(\bkappa_1^2+\bkappa_2^2)(\bq^2 +2(\bq\cdot \bkappa_1 +
\bq\cdot \bkappa_2)+4\bkappa_1\cdot \bkappa_2)$, one would find that
all the contributions are finite at $\bq^2\to 0$. For instance, in the DIS regime
of $\varepsilon^2 \gg Q_A^2$, the contribution from the term $\propto \bkappa_1\cdot \bkappa_2$
will be proportional to $\Big\{\int d^2\bkappa \bkappa_j \Phi(\bb,x_\Pom,\bkappa)
\left[\bfpsi_i(\mu^2,\bq)- \bfpsi_i(Q_A^2,\bq+\bkappa)\right]\Big\}^2$. The 
explicit forms of $\Omega(\bq^2)$ and $\Phi(\bb,x,\bkappa)$  allow a comprehensive 
study of the 
interplay of three scales -- $\bq^2,Q_A^2$ and $\varepsilon^2$ -- and we obtain 
a simple formula, which interpolates from real photoproduction to DIS and from
soft to hard gluons:
%% \bea
%% {\partial \over \partial \log{1\over \beta}}{d\sigma_A^D \over dz d^2\bb d^2\bq } \sim 
%% {\frac{1}{ \pi^3}} {\alpha_S C_FP_{q\gamma}(z)
%%  \over \bq^2 + 2\varepsilon^2 +Q_A^2}\left({Q_A^2\over Q_A^2 + \bq^2}
%% \right)^2 \cdot  \left[{\alpha_S(\bq^2+Q_A^2)G_N(x,\bq^2+Q_A^2)
%% \over \alpha_S(Q_A^2)G_N(x,Q_A^2)}\right]^2,
%% \label{eq:25}
%% \eea
\bea
&&{\partial \over \partial \log{1\over \beta}}{d\sigma_A^D \over  d^2\bb d^2\bq } \sim 
{\frac{1}{ \pi^3}} \alpha_S \alpha_{em} e_f^2 C_F N_c\nonumber\\
&\times&
\left({Q_A^2\over Q_A^2 + \bq^2}
\right)^2 \cdot \left[{\alpha_S(\bq^2+Q_A^2)G_N(x,\bq^2+Q_A^2)
\over \alpha_S(Q_A^2)G_N(x,Q_A^2)}\right]^2
\cdot {1 \over Q^2}\log {\bq^2 + Q^2 +Q_A^2 \over \bq^2 +Q_A^2}.
\label{eq:26}
\eea
As far as the dependence on three scales is concerned, 
Eq. (\ref{eq:26}) would hold for the free-nucleon
target too. Notice, that compared to inclusive DIS, 
the scaling violation in (\ref{eq:26}) is short of
one power of $\log Q^2$, as it was shown already in
\cite{NZ94,NSZZval}.
The gluon-jet spectrum has (i) a plateau in the soft region, $\bq^2\lsim Q_A^2$, 
(ii) the
$1/\bq^4$ behavior for intermediate momenta $Q_A^2\lsim \bq^2\lsim Q^2$
(for free nucleons this result is known for quite a time
\cite{KolyaMC,NSZZval}) and (iii) the $1/\bq^6$ asymptotics for $\bq^2 \gsim Q^2$.
The latter regime can be viewed as the diffraction dissociation, $q \to qg$, of the isolated
quark from the $q\bar{q}$ Fock state of the photon, in the dipole space it corresponds
to rare $q\bar{q}g$ configurations with $\brho^2 \ll \br^2$, while the bulk of DDIS comes
from $\brho^2\gsim \br^2$ \cite{NZ94,GNZ95}. The results for nuclear mass number dependence are
instructive: in hard regions (ii) and (iii) the impact parameter integration 
gives $d\sigma_A^D \propto \int d^2\bb \, \,  Q_A^4(\bb,x_\Pom) \sim A^{4/3}$, while in the plateau
region $ d\sigma_A^D \propto \int d^2\bb \, \,  Q_A^{-2}(\bb,x_\Pom)\sim A^{1/3}$ for $Q^2 \ll Q_A^2$ and  
$ d\sigma_A^D \propto \int d^2\bb \sim A^{2/3}$ for $Q^2 \gsim Q_A^2$. The width of the plateau has only
marginal dependence on $Q^2$, the  quantity
$\bq^2  d\sigma_A^D$ will take its maximum value at $\bq^2 \approx Q_A^2$ independent of
$Q^2$. Prior to our analytic formula (\ref{eq:26}), the $Q_A$ and $Q$ dependence
of the spectrum of gluons was studied numerically \cite{Golec} within 
a model which corresponds to a Gaussian parameterization for $\phi(\bb,x,\bkappa)$.
The numerical trends observed in  \cite{Golec} are consistent with our Eq. (\ref{eq:26}).

We come to a summary. We derived the nonlinear $k_\perp$-factorization formulas,
and reported full analytic results, for the inclusive spectrum of
leading quarks and gluons in the photon and pomeron fragmentation
regions of high-mass DDIS, respectively. We have also 
derived the nonlinear $k_\perp$-factorization representation for the unintegrated
glue in the pomeron as defined by the LL${1\over \beta}$ evolution of DDIS cross section in the color 
dipole representation. Our main conclusion is that, in the general case, the LL${1\over \beta}$ 
evolution-defined glue does not have the unitarity properties appropriate for a description of the jet 
spectra and DDIS final states. This mismatch entails a breaking of the familiar collinear
factorization treatment of final states and is especially acute for nuclear targets with
large saturation scale, but it would become relevant also to DDIS off free-nucleon
targets at very small $x_\Pom$, when the saturation effects become substantial 
for DIS off free nucleons too. The issue of numerical corrections to the DLLA 
for DDIS off free nucleons 
needs further scrutiny in view of the improved statistics at HERA.

This work has been partly supported by 
the DFG grant 436RUS17/138/05.

\end{document}